\documentclass[intlimits,twoside,a4paper]{article}
\usepackage{amsmath,amssymb}
\usepackage{graphicx}
\usepackage{wrapfig}

\usepackage[T2A]{fontenc}
\usepackage[cp1251]{inputenc}

\usepackage[eqsecnum]{cmpj}
\issue{2011}{14}{2}{23005}

\doinumber{10.5488/CMP.14.23005}

\title[Non-equilibrium reversible dynamics of work production...]
{Non-equilibrium reversible dynamics of work production
 in  four-spin system in a magnetic field}
\author[E.A. Ivanchenko]{E.A. Ivanchenko\thanks{E-mail: yevgeny@kipt.kharkov.ua}}
\address{Institute for Theoretical Physics,
National Science Center ``Kharkov Institute of Physics and Technology'',  1~Akademichna~Str., 61108~Kharkiv, Ukraine}

\date{Received January 5, 2011, in final form May 25, 2011}

\authorcopyright{E.A. Ivanchenko, 2011}

\begin{document}

\maketitle

\begin{abstract}
A closed system of the equations for the local Bloch vectors and
spin correlation functions is obtained by decomplexification of
the Liouville-von Neumann equation for 4 magnetic particles with
the exchange interaction that takes  place in an arbitrary
time-dependent external magnetic field.  The analytical and
numerical analysis of the quantum thermodynamic variables  is
carried out depending on separable mixed initial state and the
magnetic field modulation. Under unitary evolution, non-equilibrium
reversible dynamics of power production  in the finite environment
is investigated.
\end{abstract}

\section{Introduction}

 The classical thermodynamic heat engine converts heat energy
into mechanical work with the help of a classical mechanical
system in which some gas expands and pushes a piston in a
cylinder. Such heat engine receives  energy from a
high-temperature reservoir. Part of energy from this
reservoir is converted to mechanical work, and part of it is
transferred to a low-temperature reservoir. The classical heat
engine reaches its peak efficiency when it is reversible. Due to
the impossibility of constructing an ideally reversible heat
engine, in 1824 Carnot~\cite{Carnot} offered the mathematical
model for an ideal heat engine which is not only reversible, but
is also cyclic. In the last decades  great efforts  have been
made to investigate  the quantum properties of the
working substance, the search and practical implementation of the
quantum analogue
of the Carnot  cycle  in microsystems.

 The operation of quantum heat engines that employ
multi-level systems as working agents, for example harmonic
oscillators, free particles in a box, three-level atoms or
electrons subjected to magnetic fields, were introduced  in~\cite
{Scovill,GevaKosloff,FeldmannGevaKosloff}. The two-level quantum
systems similar to particles with spin 1/2 are essential
ingredients for quantum computation, but the coupled spin systems
can also be used as quantum  thermodynamic engines~\cite{Feldmann
Kosloff, Tova Feldmann and Ronnie Kosloff, RezekKosloff,
TonnerMahler, HenrichMahlerMichel, AllahverdyanJohalMahler}. The
quantum analogue of the Carnot cycle requires a dynamical
description of the working medium, the power output and the heat
transport mechanism. The  spin system ``working gas'' in an
external field has its own physical properties. These properties at weak interaction
with the  environment (heat baths) can, as a rule, be deformed
slightly. The purpose of this paper is to predict these properties
and find out what can be expected concerning the derivation of an
equation for the reduced matrix of a system in contact with an
environment, usually considered in the Markovian approximation.

Based on quantum mechanics, we investigate the
limiting case when interactions of the working gas, consisting of
4 particles with spin l/2 in a variable magnetic field, with heat
baths are to equal to zero. In the other words, a unitary time
evolution is studied, and the state of the system during this
evolution is interpreted using thermodynamic  concepts.

The paper is organized as follows. In  section~\ref{Hamiltonian}
we introduce the model Hamiltonian. In
section~\ref{Decomplexification} we use the Bloch representation,
in terms of the local Bloch vectors and spin correlation
functions, to write down the Liouville-von Neumann equation for a
density matrix for four particles with spin 1/2 with an exchange
interaction in a variable magnetic field. We describe the
conservation laws which effectively supervise the numerical
calculations. In section~\ref{thermodynamic variables} we describe
the local quantum thermodynamic parameters of the spin subsystems.
Our numerical results are detailed  in section~\ref{Numerical
results}. For the separable mixed initial state we  numerically
investigate the quantum thermodynamics of  a particle in the
environment of three others depending on the modulation of an
operating field and initial states. Our results are summarized and
discussed in section~\ref{Summary}. It will be numerically found
that under unitary dynamics,  the work production of one part of a
system is compensated by absorption of work  produced by the other
part. Work  production by a subsystem is accompanied by
entropy growth and vice-versa. The entropy decreases when work is
absorbed by the subsystem. Some necessary additional details for
numerical results are presented in the appendix.

 \section{Model Hamiltonian}\label{Hamiltonian}

The Hamiltonian of four coupled particles $e,\ p,\ n,\ u $ with
spin $ 1/2$ in the external ac magnetic field
$\textbf{h}=(h_1\,,h_2\,,h_3)$ looks like
\begin{eqnarray}\label{eq:1}
\hat{H} &=h^{\re}_is^{\re}_i+h^{\rm p}_is^{\rm p}_i
+h^{\rm n}_is^{\rm n}_i+h^{\rm u}_is^{\rm u}_i+2(J^{\rm ep}s^{\re}_is^{\rm p}_i
+J^{\rm en}s^{\re}_is^{\rm n}_i+J^{\rm eu}s^{\re}_is^{\rm u}_i\nonumber\\
 &\quad +J^{\rm pn}s^{\rm p}_is^{\rm n}_i+J^{\rm pu}s^{\rm p}_is{\rm ^u}_i+J^{\rm nu}
 s^{\rm n}_is{\rm ^u}_i),
\end{eqnarray}
where $h^{\re}_i\,, \ h^{\rm p}_i\, ,\ h^{\rm n}_i\,, \ h^{\rm
u}_i$ are the Cartesian components of the external magnetic field
in the energy units, operating on the corresponding particle ( we
set the Bohr magneton $\mu_{\rm B} $ equal to 1);
 $s^{\re}_i=\frac{1}{2}\sigma_i\otimes\sigma_0\otimes\sigma_0\otimes\sigma_0$,
 $s^{\rm p}_i=\frac{1}{2}\sigma_0\otimes\sigma_i\otimes\sigma_0\otimes\sigma_0 $,
  $s^{\rm n}_i=\frac{1}{2} \sigma_0\otimes\sigma_0\otimes\sigma_i\otimes\sigma_0 $,
 $s^{\rm u}_i=\frac{1}{2}\sigma_0\otimes\sigma_0\otimes\sigma_0\otimes\sigma_i$
     are  the   matrix representation of spin operators \eqref {eq:1};
    the Pauli matrices are
\[ \sigma_0=\left(
\begin{array}{cc}
  1 & 0 \\
  0 & 1 \\
\end{array}
\right), \qquad \sigma_1=\left(
\begin{array}{cc}
  0 & 1 \\
  1 & 0 \\
\end{array}
\right), \qquad \sigma_2=\left(
\begin{array}{cc}
  0 & -i \\
  i & 0 \\
\end{array}
\right), \qquad \sigma_3=\left(
\begin{array}{cc}
  1 & 0 \\
  0 & -1 \\
\end{array}
\right);\]
$\otimes$ is the symbol of direct product \cite{Lankaster};
$J^{\rm ep},\ J^{\rm en},\ J^{\rm eu},\ J^{\rm pn},\ J^{\rm pu},\
J^{\rm nu}$ are the constants of isotropic exchange interaction
between spins; the summation over  $e,\ p,\ n,\ u $ is absent.

\section{Decomplexification of the Liouville-von Neumann
equation}\label{Decomplexification}

 The Liouville-von Neumann
equation for the density matrix  $ \rho $, describing the dynamics
of  four-spin system, looks like
\begin{equation}\label{eq:2}
\ri\partial_t\rho=[\hat{H},\rho],\qquad \rho(t=0)=\rho_0\,.
\end{equation}
Let us present the solution of the equation \eqref{eq:2} as
\begin{equation}\label{eq:3}
  \rho=\frac{1}{16}R_{\alpha\beta\gamma\delta}
  \sigma_\alpha\otimes\sigma_\beta\otimes\sigma_\gamma\otimes\sigma_\delta\,,\qquad
  \rho^+=\rho, \qquad {\rm Tr}\rho=1, \qquad R_{0000}=1.
\end{equation}
 Hereinafter summation is taken over  the repeating  Greek indices
 from zero up to three,
and over Latin indices from one up to three. The four coherence
vectors  (the Bloch vectors)    widely used in  the magnetic
resonance theory, are written as
\begin{subequations}\label{eq:4}
\begin{equation}\label{eq:4a}
R_{i000}={\rm
Tr\,}\rho\,\sigma_i\otimes\sigma_0\otimes\sigma_0\otimes\sigma_0\,,
\end{equation}
\begin{equation}\label{eq:4b}
R_{0i00}={\rm
Tr\,}\rho\,\sigma_0\otimes\sigma_i\otimes\sigma_0\otimes\sigma_0\,,
\end{equation}
\begin{equation}\label{eq:4c}
R_{00i0}={\rm
Tr\,}\rho\,\sigma_0\otimes\sigma_0\otimes\sigma_i\otimes\sigma_0\,,
\end{equation}
\begin{equation}\label{eq:4d}
R_{000i}={\rm
Tr\,}\rho\,\sigma_0\otimes\sigma_0\otimes\sigma_0\otimes\sigma_i\,.
\end{equation}
\end{subequations}
 These vectors  characterize the local properties of
individual spins, whereas the other tensors describe the spin
correlations. All correlation functions are in
   the limits
   \begin{equation}\label{eq:5}
   -1 \leqslant R _ {\alpha\beta\gamma\delta} \leqslant 1.
   \end{equation}
 As $ \ri\partial_t\rho^n = [\hat {H}, \rho^n]$,
$ (n=1,\ 2,\ 3, \dots) $ at unitary evolution there is a enumerable
number of conservation laws $ {\rm Tr \,} \rho=C_1=1, ~ {\rm Tr
\,} \rho^2=C_2\,, \dots $, where $C_n  $ are the  constants of
motion, from which only the first $C_2\,,\ C_3\,,\ldots,\ C_{16} $
are algebraically independent~\cite {Tapia}. From the conservation
of purity, for which $ (\rho^2) _ {ik} \stackrel {\rm def} \equiv
(\rho) _ {ik} $\,, the polynomial (square-law) invariants  are
obtained. The square polynomials also  control  the signs  $R _
{\alpha\beta\gamma\delta}$\,.
 The length of the generalized
Bloch vector $ b^{\rm epnu} $   is conserved under unitary
evolution:
\begin{equation}\label{eq:6}
b^{\rm epnu} =\sqrt{R^2_{\alpha\beta\gamma\delta}-1}\,.
\end{equation}
Having inserted the equation \eqref{eq:3} into \eqref{eq:2},
multiply the equation \eqref{eq:2} by all elements of the basis
$\sigma_\alpha\otimes\sigma_\beta\otimes\sigma_\gamma\otimes\sigma_\delta$
 in turn and take the trace for each equation to get the time derivatives
 of the correlation functions as $R_{\alpha\beta\gamma\delta}$
\begin{equation}\label{eq:7}
\ri\partial_t R_{\alpha\beta\gamma\delta}={\rm Tr\,}
[\hat{H},\rho]
\sigma_\alpha\otimes\sigma_\beta\otimes\sigma_\gamma\otimes\sigma_\delta\,
; \qquad \alpha=(0,\ 1,\ 2,\ 3),\dots, \qquad \delta=(0,\ 1,\ 2,\
3).
\end{equation}
A detailed form of the system equation \eqref{eq:7} is in the
appendix. The derivation algorithm of the system
equation~\eqref{eq:A.1}--\eqref{eq:A.15} is presented
in~\cite{Ivanchenkojmp}.  The Liouville-von Neumann equation
accepts a real form in terms of the functions $R _
{\alpha\beta\gamma\delta} $ as a closed system of 255 differential
equations  for the local Bloch vectors and spin correlation
functions.

In case of equivalent particles at $ \mathbf{h}^{\rm e} =\mathbf{h}^{\rm p}
=\mathbf{h}^{\rm n} =\mathbf{h}^{\rm u}=\mathbf{h},\; J^{\rm
ep}=J^{\rm en}=J^{\rm eu}=J^{\rm pn}=J^{\rm pu}=J^{\rm nu}=J $
from the equations~\eqref{eq:A.1}--\eqref{eq:A.4}
and~\eqref{eq:A.5}--\eqref{eq:A.10} it follows that   the  square
length of the total magnetization $ (R _ {q000} +R _ {0q00} +R _
{00q0} +R _ {000q}) ^2$ and the forms $R _ {ii00}\,,\ R _
{i0i0}\,,
 \ \ldots, R _ {00ii}$ are conserved.

In the system~\eqref{eq:A.1}--\eqref{eq:A.15}, assuming, for
example,
    $J ^ {\rm eu}=J ^ {\rm pu}=J ^ {\rm nu} =0 $,
 we get a closed system of  equations for the description of
three-qubit dynamics~\cite{Ivanchenko}.

The set of equations~\eqref{eq:A.1}--\eqref{eq:A.15} with the given
initial  conditions  has  wide applications, since the magnetic
field enters the form of  arbitrary functions. First of all, it allows to
make  numerical calculations  for continuous (paramagnetic resonance in a continuous mode) as well as
for pulse modes (nuclear magnetic resonance). Secondly,
by means of this system it is possible to investigate the entanglement dynamics
of qubits in a magnetic field~\cite{VerstraeteDehaeneDe MoorVerschelde}
since the entanglement measures are expressed in terms of the reduced density matrices or  of populations.
Thirdly, an important application  of the system~\eqref{eq:A.1}--\eqref{eq:A.15}
  is quantum approach to  the Carnot cycle~\cite{Carnot, Scovill, GevaKosloff, FeldmannGevaKosloff, Feldmann
Kosloff, Tova Feldmann and Ronnie Kosloff, RezekKosloff}, where the
working body is a finite spin chain.

\section{Quantum thermodynamic variables}\label{thermodynamic variables}

 In the external  field, the energy of a four-spin system is defined by the formula
\begin{eqnarray}\label{eq:8}
E&=&{\rm Tr\,}\hat{H}(t)\rho
  =\frac{1}{2}\left(h_i^{\rm e}R_{i000}+h_i^{\rm p}R_{0i00}+h_i^{\rm n}R_{00i0}+h_i^{\rm u}
  R_{000i}\right)\nonumber\\
  &&{}+2(J^{\rm ep}R_{ii00}+J^{\rm en}R_{i0i0}+J^{\rm pn}R_{0ii0}+
 J^{\rm eu}R_{i00i}+J^{\rm pu}R_{0i0i}+J^{\rm nu}R_{00ii}).
\end{eqnarray}
The change of the total energy expectation value is equal to
$\partial_t E=\partial_t {\rm Tr\,}\hat{H(t)}\rho={\rm
Tr\,}\partial_t\hat{H(t)}\rho +{\rm Tr\,}
\hat{H(t)}\partial_t\rho$. Due to the equation of motion~\eqref
{eq:2}
 we have ${\rm Tr\,}\hat{H(t)}\partial_t\rho=-\ri{\rm Tr\,} \hat{H(t)}[ \hat{H(t)},\rho]=0$.
For an external static field,
the Hamiltonian  is independent of time, hence $\partial_tE=0$, that
is the energy of the system is constant in
time.

 The change of work $W$ can be associated with a term
where only the spectrum changes $
\partial_tW= {\rm Tr\,}\partial_t\hat{H(t)}\rho$. The change of
heat $Q$ is then $ \partial_tQ={\rm
Tr\,}\hat{H(t)}\partial_t\rho$. We have,  $\int _ {0} ^ {t}\rd
t\partial_t W =\int _ {0} ^ {t}\rd t {\rm
Tr\,}\partial_t\hat{H(t)}\rho$ = (integration by parts)= ${\rm
Tr\,}\hat{H(t)}\rho(t)-{\rm Tr\,}\hat{H(0)}\rho(0)- \int _ {0} ^
{t}\rd t  {\rm Tr\,} [\hat{H(t)},\partial_t\rho]= \Delta W_{\rm
sys}$\,. The last integral is equal to zero due to the equation of
motion. Therefore,  the work production in the system is given by an expression of the form:
\begin{eqnarray}\label{eq:9}
\Delta W _ {\rm sys} &=& {\rm Tr \,} \hat{H} (t) \rho (t) - {\rm
Tr \,} \hat{H} (0) \rho_0 \nonumber\\&=&\int _ {0} ^
{t}\frac{1}{2}\left[ (\partial_th_i^{\rm e})R_{i000}+
(\partial_th_i^{\rm p})R_{0i00}+ (\partial_th_i^{\rm n})R_{00i0}+
(\partial_th_i^{\rm u})R_{000i} \right]\rd t.
\end{eqnarray}
 The change in finite time $t$ of the full system energy which, in our problem,
consists of the performed work and heat energy is equal to the work production since in the closed system heat energy
is not produced.

 The reduced density matrices  describe the
dynamics of subsystems (4 matrices of individual particles, 6
matrices of two particles, 4 matrices of three particles) and, for
example, for a particle $e $, and
 the coupled particles $ep $ and $epn $, it can be written as
 \begin{subequations}
\begin{equation}\label{eq:10a}
  \rho^{\rm e}(t)={\rm Tr\,}_{\rm pnu}\rho=\frac{1}{2}\left(%
\begin{array}{cc}
  1+R_{3000} & R_{1000}-\ri R_{2000} \\
  R_{1000}+\ri R_{2000}(t) & 1-R_{3000} \\
\end{array}%
\right),
\end{equation}
\begin{equation}\label{eq:10b}
  \rho^{\rm pn}=\frac{1}{4}R_{0\beta\gamma0}
  \sigma_\beta \otimes\sigma_\gamma\,, \qquad
  \rho^{\rm epn}=\frac{1}{8}R_{\alpha\beta\gamma0}
  \sigma_\alpha\otimes\sigma_\beta \otimes \sigma_\gamma\,.
\end{equation}
\end{subequations}
 The matrices~\eqref{eq:10a}--\eqref{eq:10b} are determined by the system
solution~\eqref{eq:A.1}--\eqref{eq:A.15}, since the equations for
the reduced matrices are not closed.  From the
system~\eqref{eq:A.1}--\eqref{eq:A.15} it
 follows that the spin flip probabilities of  $p,\,n $\, from their
initial state  are equal to
\begin{equation}\label {eq:11}
  P^{\rm p} =\frac {1-R _ {0300}} {2}\, ,\qquad  P^{\rm n} =\frac {1-R _ {0030}} {2}\,.
\end{equation}

In the longitudinal field  $ \textbf {h} = (0,0, h_3) $ with equal
coupling constants
 $J $ on the equations~\eqref{eq:A.1}--\eqref{eq:A.4}
 it follows that
\begin{equation}\label {eq:12}
\partial_t m_q = \varepsilon _ {3sq} h_3 m_s\,,
\end{equation}
where $ m_q=R _ {q000} +R _ {0q00} +R _ {00q0} +R _ {000q} $ and $
m_{3} $   is the invariant of motion. The system of equations for
$m_1\,, \ m_2$ has  zero solutions $m_1=0, \; m_2=0$ for the
initial condition~\eqref {A.16}. We believe that transverse
coherences $R_{q000} =R _ {0q00} =R _ {00q0} =R _ {000q}=0$ \; for
$q=1,\ 2 $ because the numerical solution confirms that each term
 in $m_1\,, \ m_2$  is equal to zero during unitary evolution,
that is the reduced matrix
\[
\rho^{\rm e}=\frac{1}{2}\left(%
\begin{array}{cc}
  1+R_{3000} & 0 \\
  0 & 1-R_{3000} \\
\end{array}%
\right) \] is diagonal. In this case it is possible  to correctly define
the local dynamic temperature.  The local or dynamic
temperature for a two-level system can be defined according to the
basic meaning of a thermal
state~\cite{HenrichMahlerMichel,WeimerMahler}:
\begin {equation}\label{eq:13}
T^{\rm e}(t) =-\frac {\Omega (t)} {k_{\rm B}\ln ( {p_1}/
{p_0})}\,,
\end {equation}
where  $ \Omega (t) $ is the transition frequency in the $e $
two-level  system, equal to $  h_3 $\,. We  assume that the
interaction strength  J  between the particles is small in
comparison to the energy level spacing. In this case,  the
interaction will not significantly alter the instantaneous  energy
eigenvalues of the system  and hence we can meaningfully define
the temperature of individual particles, since each will
remain in the  instantaneous  thermal form~\eqref{eq:13}, with the
energy level spacing being the same as in the absence of interaction.
Further we set the Bolzmann constant $k_{\rm B} $ equal to 1.
Therefore, all temperatures are in energy units.  It is necessary
to note a generalized approach to temperature, work and heat
without weak coupling approximation between the
particles~\cite{WeimerHenrichRemppSchroderMahler}.

 The $e$-spin  entropy is equal to
\begin{equation} \label {eq:14}
  S^{\rm e}(t) = - {\rm Tr \,}\rho^{\rm e} \ln
  \rho^{\rm e} =-p_0 \ln p_0-p_1\ln p_1\,,
\end{equation}
  where the local populations are equal to
\begin{equation} \label {eq:15}
    p_0=\frac {1-R _ {3000}} {2}\,,\qquad p_1=\frac {1+R _ {3000}}
  {2}\,.
  \end{equation}
  As the system is isolated, its entropy as a whole is constant.
  Hence, the system is reversible.  In the reversible system, the subsystems
  should be reversible, that is, there exist entropy fluxes between subsystems
  if the initial state is non-equilibrium. The    entropy rate $\partial_t
  S^{\rm e}$
  of the $e $-spin due to the $p,\ n,\ u$ environment  is defined by the
  formula:
\begin{equation} \label{eq:16}
  \partial_t S^{\rm e}  =\frac {\partial_tR _ {3000}
  } {2} \ln\frac {1-R _ {3000}} {1+R _ {3000}}\,.
\end{equation}
The stated~\eqref{eq:13},~\eqref{eq:14} and~\eqref{eq:16} concern each particle.

 The calculation of the work carried out by subsystems
is done in the papers~\cite{HenrichMahlerMichel} by means of $ST$
diagrams. It is shown numerically that the entropy $S$ and
temperature $T$ are dependent thermodynamic variables. For a
closed trajectory in the $ST$ plane, the change of the total energy
$ \Delta W _ {\rm sys} $ is equal to zero and consequently the
area captured by the closed curve in the $ST $ plane determines
the work during a reversible cycle
\begin{equation} \label {eq:17}
 \Delta W =-\oint T \rd S =-\int _ {0} ^ {t_{\rm c}} T (t) \partial_t S\rd t,
\end{equation}
where $t_{\rm c} $ is the duration of a cycle, and  the  sign is
defined according to the rule saying that if, at a path tracing clockwise, the
area is situated on the right it obtains a minus sign (heat
 pump). The spin system is  isolated and consequently, the Carnot cycles
can only refer to sub-systems of the 4-spin system. We use~\eqref
{eq:17} for $ e,\ p,\ n,\ u$ particles.

  The energy of the \textit{coupled} particle $e$ in a magnetic field
 $h_3$
  in an environment of three others is equal to $\frac{1}{2}h_3R_{3000}$\,.
We use the formula  $
\partial_t\left(\frac{1}{2}h_3R_{3000}\right) =
\frac{1}{2}(\partial_th_3)R_{3000} +
\frac{1}{2}h_3\partial_tR_{3000}$\,. The work production (i.e., the heat
production)  by spin $e$ during a cycle is equal to
$ \int _
{0} ^ {t_{\rm c}}\frac{1}{2}(\partial_th_3)R_{3000}\rd t$ $\left(\int _ {0} ^ {t_{\rm c}}\frac{1}{2}h_3\partial_tR_{3000}\rd
t\right).$
 During that cycle the energy change of  the particle
$e$ $ \int _ {0} ^ {t_{\rm
c}}\partial_t(\frac{1}{2}h_3R_{3000})\rd t$ is equal to zero,
hence
\[
 \int _ {0} ^ {t_{\rm
c}}\frac{1}{2}(\partial_th_3)R_{3000}\rd t=- \int _ {0} ^
{t_{\rm c}}\frac{1}{2}h_3\partial_tR_{3000}\rd t.
\]
Having inserted
the equations~\eqref{eq:13}, \eqref{eq:16} into~\eqref{eq:17}
 we conclude that
\begin{equation*}
\Delta W^{\rm e} =-\oint T^{\rm e} \rd S^{\rm e}=   \int _ {0} ^
{t_{\rm c}}\frac{1}{2}(\partial_th_3)R_{3000}\rd t.
\end{equation*}
Thus, the definitions of temperature~\eqref{eq:13},  entropy~\eqref{eq:14} and work
\eqref{eq:17}
 are \textit{coordinated} with the work/heat production  for parameters
 for  which the $ST$ plots are closed. It also concerns $p,n,u$
 particles.

 The  Klein-von Neumann inequality looks like
\begin {equation} \label {eq:18}
 - {\rm Tr \,}\rho \ln \rho \leqslant-\sum _ {i=1} ^ {m} \rho _ {ii} \ln \rho _ {ii} \leqslant \ln
 m,
\end {equation}
where $-\sum _ {i=1} ^ {m} \rho _ {ii} \ln \rho _ {ii} $ is the
diagonal entropy, $d $, $m $ is the number of system states. For
the initial diagonal state, the  diagonal entropy possesses the
property $d (0) \leqslant d (t) =-\sum _ {i=1} ^ {m} \rho _ {ii}
\ln \rho _ {ii}$~\cite{BarankovPolkovnikov}.

 The dynamics of a purity measure   $P = {\rm
Tr \,} \rho {^2} $  is connected with the dynamics of entropy $S =
- {\rm Tr \,} \rho\ln\rho $ as follows. If the entropy is equal to
zero, the system is in a  pure state. At the maximum entropy, the
system is in the maximum mixed state. The purity $P $ has the
maximum value 1 for a pure state and the minimum value in the
mixed state, equal to ${1}/{m} $, where $m $ is the number of
system states. The subsystem purity is expressed in terms of the
square length of the local  or
 generalized Bloch vector~\eqref {eq:6}:
 the purity for the   subsystems $pn $ and $epnu $ is equal to
\begin{equation} \label{eq:19}
 P ^ {\rm np} = \frac {1} {4} \left(1 +
{b ^ {\rm np}} ^2\right), \qquad P ^ {\rm epnu} = \frac {1} {16}
\left(1 + {b ^ {\rm epnu}} ^2\right),
\end{equation}
accordingly, where $b ^ {np} = \sqrt {\sum _ {i, j=1} ^3R^2 _
{0ij0}} $ is the length of the generalized
Bloch vector  of the $pn $ system.

 Let us define the  entanglement measure $p $ and $n $ for spins according to~\cite{SchlienzMahler}
 on system solutions having entered
 the two-particle entanglement tensor:
\begin{equation} \label {eq:20}
 m _ {0ij0} =R _ {0ij0}-R _ {0i00} R _ {00j0}\,.
\end{equation}
The tensor $m _ {0ij0} $ is equal to zero when  the two-particle
correlation function $ R _ {0ij0} $ is  factorized in terms of the
local Bloch vectors~\eqref{eq:4a}--\eqref{eq:4d} and thus the
matrix will be separable, i.e., $ \rho ^ {\rm pn} = \rho^{\rm
p}\otimes\rho^{\rm n} $. By means of this tensor  we shall define
a measure of the two-particle entanglement in the $pn $
 subsystem
\begin{equation} \label{eq:21}
m ^ {\rm pn} _ {\rm SM} = \sqrt {\frac {1} {3} \sum _ {i, j=1} ^3m
_ {0ij0} ^2}\,.
\end{equation}
This measure is equal to zero for a separable state  and it  is
equal to $1$ for the Greenberger-Horne-Zeilinger maximum
entangled state. This measure is applicable both to the pure and
 mixed states (in all 6 two-particle measures).

\section{Numerical results}\label{Numerical results}

The quantum thermodynamic devices are subdivided into heat pumps
and heat engines depending on functional purpose. In our model, the
work absorbed or done  by the system arises due to the
displacement of the power levels  by a magnetic field~\cite{Landau,
Zener, Stueckelberg},
as well as it also depends on the initial state of the system.

 We will consider the effect of a variable magnetic field
\begin{equation}\label{eq:22}\mathbf{h} = (h_1\,,\ h_2\,,\ h_3)
\end{equation}
on the dynamics of a system in the case
when the fields operating on the spins $e$, $p$, $n$, $u$
are equal to $ \mathbf{h}  $, where $h_1=0,\, h_2=0,\,
 h_3=c +1.5 \sin \omega t $ ($c$ is the static part of the magnetic field),
with an external field frequency $ \omega=0.04$ and with all the
exchange constants equal to  0.01. The values of magnetic field frequency are
given in the energy units.

 We study the behaviour of a four-spin system in terms of local
thermodynamic concepts: temperature~\eqref{eq:13},
entropy~\eqref{eq:14} and work~\eqref{eq:17} depending on the
parameters of a possible non-equilibrium initial state:

 (i)
$e,\ p,\ u$ particles have the same temperature $T^{\rm e}(0)=
T^{\rm p}(0)= T^{\rm u}(0)=0.2$,   and the temperature of the $ n
$ particle is equal to $T^{\rm n}(0)=0.6$;

 (ii)  $T^{\rm e}(0)= T^{\rm p}(0)=0.2, T^{\rm n}(0)= T^{\rm u}(0)=0.25$;

  (iii) $T^{\rm e}(0)=0.21, T^{\rm p}(0)=0.225, T^{\rm n}(0)=0.264,
 T^{\rm u}(0)=0.28$.

\begin{figure}
% \hspace{0.1cm}
\includegraphics[width=0.48\textwidth]{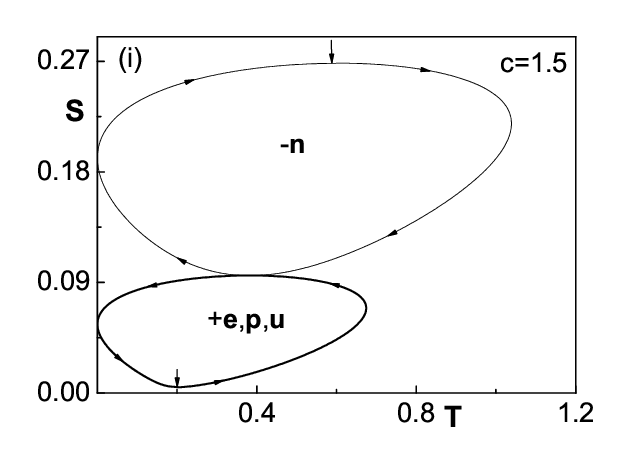}
\hfill\hspace{0.8cm}
\includegraphics[width=0.48\textwidth]{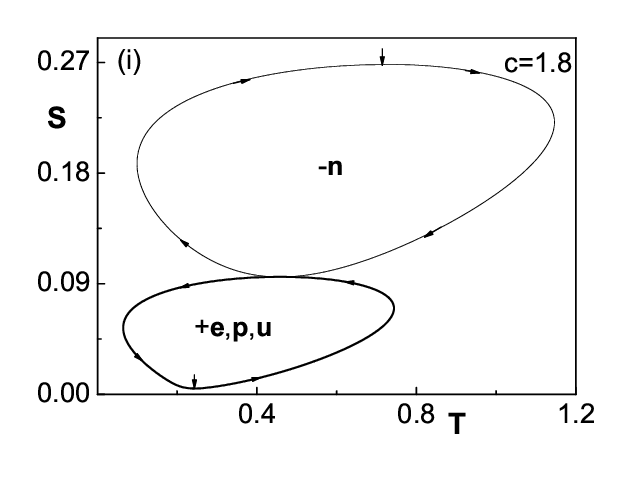} \\
  %\hfill
\includegraphics[width=0.48\textwidth]{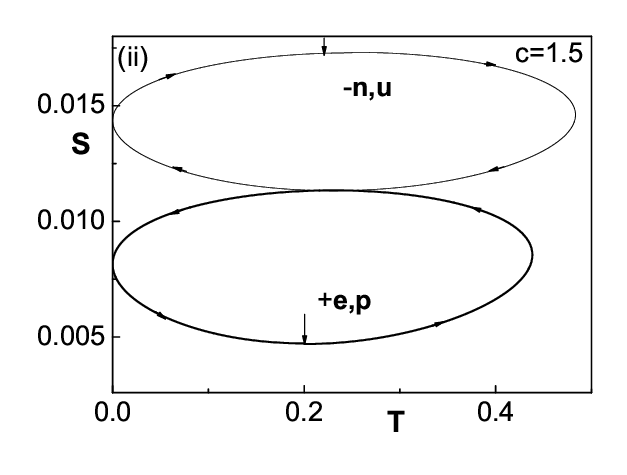}
\hfill
\includegraphics[width=0.48\textwidth]{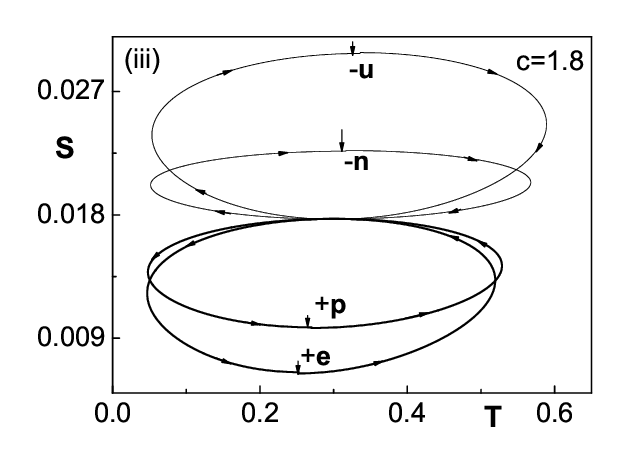}
 \caption{\label {Perfectcyclesc15i} $ST$ diagrams for
the non-equilibrium
 initial state with the different values of the static field
  $c=h^{\rm e}_3(0)=h^{\rm p}_3(0)=h^{\rm n}_3(0)={ h}^{\rm u}_3(0)$ and
 initial temperatures: (i) $ T^{\rm e} (0) =T^{\rm p} (0) =T^{\rm
 u}(0) =0.2,\ T^{\rm n} (0) =0.6$, \ (ii) $T^{\rm e}(0)= T^{\rm
p}(0)=0.2,\ T^{\rm n}(0)= T^{\rm u}(0)=0.25$, \ (iii) $T^{\rm
e}(0)=0.21,\ T^{\rm p}(0)=0.225,\ T^{\rm n}(0)=0.264,\
 T^{\rm u}(0)=0.28$. All spins are in an identical field $
\mathbf{h} $ and the exchange constants
 are equal to $0.01$. It is seen that the field $\mathbf{h} = (0, 0, c +1.5 \sin \omega
t)$ causes  periodic cycles  with parameters of the  initial state
 $\omega=0.04$. The duration of a cycle is equal to $t_{\rm c}=2 \pi/ \omega=157.08$.
The arrows indicate the  direction of circulation. The circulation
direction becomes  opposite for all particles after the replacement of
a frequency sign $ \omega $. In addition to the  direction of
circulation, the signs $ +, - $ specify the  negative or the positive
work of particles. The vertical arrows in the plots specify the
return points.}
\end{figure}

For the parameters (i) of the initial state~\eqref {A.16},
let us choose the following values for our numerical results:
$h^{\rm e}_3(0)=h^{\rm p}_3(0)=h^{\rm n}_3(0)=h^{\rm u}_3(0)=c$.
 In figure~\ref{Perfectcyclesc15i} (i), c=1.5   the parametric
dependences on entropy and temperature for a cycle are shown, i.e.,
$ST $ cycles with negative work (bold line) which are identical
for $e,\ p,\ u $ spins. The spin $n $ produces some work (thin
line) from the ($e,\ p,\ u $)~-- environment, while each of the
spins $e,\ p,\ u $ (bold line) absorbs work.
The work performed by $n $ spin, is equal to the area limited
by a closed curve according to the formula~\eqref {eq:17} and is
equal to  0.1331514. The $e,\ p,\ u $ spins perform an amount of
work equal to $3\times(-0.0443838) = - 0.1331514$. Thereby, the
work production  of the four-spin system $ \Delta W _ {\rm sys}
$~\eqref {eq:9} is equal to zero in full compliance with the
general results for the isolated system~\cite{LandauLifshitz,
Callen, Balian, EspositoMukamel,SchroderTeifelMahler}. It confirms
the use of temperature~\eqref {eq:13} and entropy~\eqref {eq:14}
as effective thermodynamic characteristics. A feature of this
initial state and modulation magnetic field is that \textit{these
cycles are repeated  without any deformation}. (It is known that
the system equations~\eqref {eq:A.1}--\eqref {eq:A.15} with
periodic coefficients, according to the Floquet theory,  have
periodic or quasi-periodic solutions  i.e. the Floquet theory does
not exclude the periodic solutions as well. This depends on the set of
coefficients. We have presented this set.) In other words, each
cycle comes to an end  returning to the same initial state. The
closure of the $ST$ plots does not depend on the amplitudes of the
driving field (only the form and the area may change), but it
critically depends on the frequency $ \omega
$~\cite{HenrichMahlerMichel} and the module of the exchange
constants, it does not depend on ferromagnetic or
antiferromagnetic character of the working gas~\cite {Guo-Feng
Zhang}. After replacement of a  frequency sign
  $ \omega $, the circulation direction becomes  opposite for all
particles. If the  value of the temperature parameter is $T^n (0)> 0.2$, then the particle $n $ does work, and the spins $e, p, u $ absorb work. The purity of the whole system decreases with an increase of $T^n (0) $, but cycle-after-cycle and
periodicity remain as it is described,  and the areas characterizing work increase
 with preservation of the algebraic sum which is equal to zero. For $T^n (0) <
 0.2$ the  spins $e, p, u $ produce work, and the spin $n $ absorbs it.

 If $T^{\rm n} (0) $ approaches 0.2, that is,  at the equilibrium
initial state $T^{\rm e} (0) = T^{\rm p} (0) = T^{\rm n} (0)
=T^{\rm u}(0) =0.2$ the entropy of each subsystem is constant and
the work production of each subsystem is equal to zero, since at
the initial moment
  there is no temperature gradient in the system
  (passive or immovable state~\cite{AllahverdyanBalianNieuwenhuizen}).

For  the  initial state~(i), c=1.8 the work performed by the $n $
spin   is equal to $ 0.1329435$. The $e,\ p,\ u $ spins perform an
amount of work equal to $3 \times ( -0.0443145) = - 0.1329435$.

The numerical calculations also show that for  the  initial
  state~(ii), c=1.5    particles $e,\ p$ absorb work equal to $2( - 0.002265)$,
  and particles $n,\ u$ produce  the same work $2( + 0.002265)$~\cite{Boykin}.

In the case of initial state~(iii) for c=1.8 particles $e,\
p$ absorb work which is  equal to  $(-0.0041)+(-0.0029)$,
  and particles $n,\ u$ produce  the same work  $(0.0020+0.0050)$.

We  have the results of the $ST$ diagrams in
figure~\ref{Perfectcyclesc15i} at different static parts of the
magnetic field $c$ for the  initial non-equilibrium states. The
numerical analysis shows that the balance of   work production for
(i), (ii), (iii) in fact does not depend on the static magnetic
field, only the   local temperatures and  the form of the $ST$
diagrams  change.

In figure~\ref{Perfectcyclesc15i} it is seen that in the middle
of    the cycle, the non-equilibrium reversible system
   is converted to the quasi-equilibrium one;
      the entropy rates $\partial_t S^i$    change signs,
    the temperature rates $\partial_t T^i$  are minimum
    $(i=e,\ p,\ n,\ u)$.

We would like to indicate that in the vicinity of $t\approx117.8$, the
local temperature of all particles goes  to
 zero as the frequencies $ \Omega^i (t) $ for all particles go
to zero.  It is necessary to notice that for the minimum local
temperatures, the eigenvalues of the Hamiltonian~\eqref {eq:1} come
closer to zero, and with the   growth of   temperatures the
eigenvalues  become bigger. The transition probability of each
particle~\eqref{eq:11} is close to 1 and makes one oscillation per
cycle. The  population  $ \rho_{16 \, 16}
 $   is approximately equal to 0.925 during a cycle. For
the opposite sign $h^{\rm e}_3(0)=h^{\rm p}_3(0)=h^{\rm
n}_3(0)=h^{\rm u}_3(0) =-1.5,\, \rho_{1 \, 1}  \approx 0.925$ the
thermodynamic characteristics do not change. In figure~\ref{Perfectcyclesc15i} it is clearly seen that if the initial
state of a subsystem is more disordered, a subsystem
absorbs/produces   more (in absolute magnitude)
work~\cite{AllahverdyanBalianNieuwenhuizen}.

 The cycle of each particle is determined by the
direct coupling
 of the given particle with the others  and does not depend on the coupling constants
between other particles.  The calculations confirm that the
dependence between   the energy of particles $p,\ n$ $E ^ {\rm pn}
=  (h^{\rm p}_iR _ {0i00} +h^{\rm n}_iR _ {00i0})/2 +2J ^ {\rm pn}
R _ {0ii0} $ and the entropy $S ^ {\rm pn} $ is also cyclic. This
property is carried out by any particle pair.

 \begin{figure}[ht]
\includegraphics[width=0.48\textwidth]{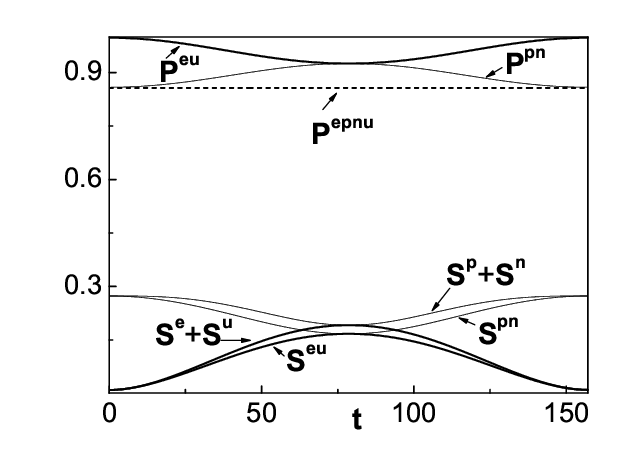}
 \hfill
\includegraphics [width=0.47\textwidth]{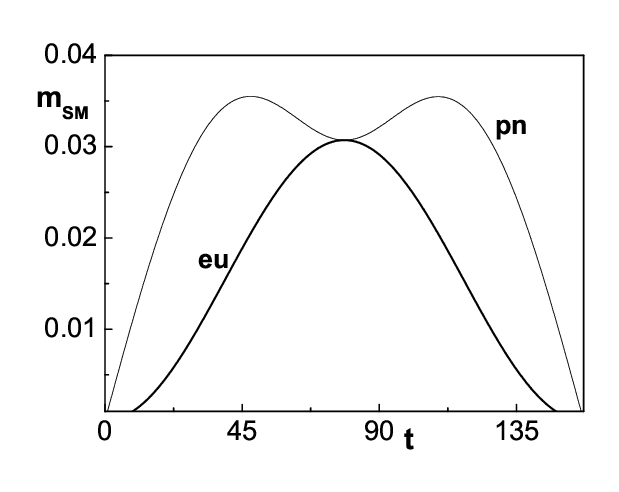}\hspace{0.4cm}
 \\
 \parbox[t]{0.5\textwidth}{\caption { Time dependence on  purity
 of all system $  {P} ^ {\rm {epnu}} $ and on characteristics $eu $ and $pn $
 subsystems for one cycle with the parameters, corresponding
 to the $ST $ diagrams in figure~\ref{Perfectcyclesc15i} (i), c=1.5.}\label {eupnProperties}}
\hfill
 \parbox[t]{0.5\textwidth}
 {\caption {  Time dependence on  the local
 entanglement $m _ {\rm SM} $ for subsystems
 $pn$ (thin line) and $eu$ (bold line)
  for one cycle. The parameters correspond to the
 $ST$ diagrams in figure~\ref{Perfectcyclesc15i} (i), c=1.5.}\label{Entangpneu}}
 \end{figure}
 The results for additional quantum thermodynamic
characteristics of $eu $ and $pn $ subsystems are presented in
figures~\ref{eupnProperties} and~\ref{Entangpneu}. Under unitary
evolution, the  global purity $P ^ {\rm epnu} $  does not depend on
time. The purity $P ^ {\rm pn} $~\eqref {eq:19}
  has a maximum at an entropy minimum. The work
production of the $eu $ subsystem  is accompanied by entropy
increase and purity reduction
 and the inverse process occurs at work absorption.
 It is seen that the local
entanglement $m ^ {\rm pn} $~\eqref {eq:21} has a maximum for a
minimum entropy $S ^ {\rm pn} = - {\rm Tr } \rho ^ {\rm pn} \ln
  \rho ^ {\rm pn} $.  The entanglement~\eqref {eq:21} between the  particles is
noticeable in the middle of the cycle and it is bigger between particles
with opposite signs in work, and, as the calculations confirm, grows
with the disorder increase in the system. This entropy is always less or  equal
to the sum entropies of  individual  spins
  in the $pn $ subsystem. Also, the inequality of Klein-von Neumann is carried
  out~\eqref {eq:18}, i. e., the diagonal entropy is bigger or  equal
  to the $pn $  subsystem entropy.
   For the case presented   in
figure~\ref{Perfectcyclesc15i} (i),  c=1.5
   the diagonal entropy $d ^ {\rm pn} $ ($d ^ {\rm eu} $) coincides with the summed
   entropies of  the individual  spins $S^{\rm p}+S^{\rm n} $ ($S^{\rm e}+S^{\rm u} $).

The control of calculations was carried out with the help of
the invariants of the motion,
       described in sections~\ref{Decomplexification} and~\ref{thermodynamic variables},
       and all the correlation functions have been  in
   the limits $-1 \leqslant R_{\alpha\beta\gamma\delta} \leqslant 1$.

\section{Summary}\label{Summary}
 A closed system of equations  is derived for the local Bloch vectors  and spin correlation functions
  of four two-level systems with exchange interaction, being in a
  time-dependent  external magnetic field. The invariants of motion necessary for the control of computing have been found.
Analytical and numerical analysis of thermodynamic behaviour in a four-spin weak coupling  system depending on
   the parameters characterizing the initial  non-equilibrium state and modulation of
   the driving  field was performed.
   It was numerically found that under unitary dynamics,  the work production of one part
   of the system is compensated by absorption of work  produced by the other part.
     The work  production by  a  subsystem is accompanied
   by entropy growth and vice-versa, while  the entropy decreases with work absorption
      by the  subsystem.

It is shown that in the middle of a cycle,
      the non-equilibrium reversible system  is converted to
       a quasi-equilibrium one. Thermalization of non-equilibrium
system of 4 spins in finite time under different initial
conditions is numerically shown (figure~1). In other words, the
finite non-equilibrium system  in  finite time  generates an
``attractor'', i.e., an intermediate temperature is established for
all particles.
       It was revealed that the $ST$ cycle of each particle
    is determined by the direct
coupling of the given particle with the others and weakly depends
on the coupling constants between the other spins.

 It was
analytically and numerically shown that  the formulas for
temperature~\eqref {eq:13} and  entropy~\eqref {eq:14} are the
effective thermodynamic characteristics for work calculation in
a weakly coupled spin  system by means of $ST$ diagrams in the
course of unitary evolution.

 The study  of the ``spin gas''
properties  is necessary for the implementation of quantum
thermodynamic   cycles in
  spin systems~\cite{Quan,He,LindenPopescuSkrzypczyk}.  Taking into account the environment,
  zero balance will be broken and the system will become a heat engine
   or a heat pump depending on the temperatures of the heat baths.

\section*{Acknowledgements}
The author is grateful to Zippa~A.A.  for constant invaluable
support and to  the referees for their objective comments that
improved the text in many points.

\newpage
\appendix

\section*{Appendix}\label{A}
\renewcommand{\theequation}{A\arabic{equation}}
The detailed form of the system equation~\eqref{eq:7} is as follows:
\begin{eqnarray}
\partial_tR_{q000}&=&\varepsilon_{isq}h^{\re}_iR_{s000}
+\varepsilon_{stq}(J^{\rm ep}R_{ts00}+J^{\rm en}R_{t0s0}+J^{\rm
eu}R_{t00s})\label{eq:A.1},
\\[1.5ex]
  \partial_tR_{0q00}&=&\varepsilon_{isq}h^{\rm p}_iR_{0s00}
  +\varepsilon_{stq}(J^{\rm ep}R_{st00}+J^{\rm pn}R_{0ts0}+J^{\rm pu}R_{0t0s}),\label{eq:A.2}
\\[1.5ex]
\partial_tR_{00q0}&=&\varepsilon_{isq}h^{\rm n}_iR_{00s0}
+\varepsilon_{stq}(J^{\rm en}R_{s0t0}+J^{\rm pn}R_{0st0}+J^{\rm
nu}R_{00ts}),\label{eq:A.3}
\\[1.5ex]
 \partial_tR_{000q}&=&\varepsilon_{isq}h^{\rm u}_iR_{000s}+\varepsilon_{stq}
 (J^{\rm eu}R_{s00t}+J^{\rm pu}R_{0s0t}+J^{\rm nu}R_{00st}),\label{eq:A.4}
\\[1.5ex]
\partial_tR_{qk00}&=&\varepsilon_{isq}h^{\re}_iR_{sk00}
+\varepsilon_{isk}h^{\rm p}_iR_{qs00}+\label{eq:A.5}
 J^{\rm ep}\varepsilon_{ksq}(R_{s000}-R_{0s00})\nonumber \\
 &&{}+\varepsilon_{tsq}(J^{\rm en}R_{skt0}+J^{\rm eu}R_{sk0t})+
 \varepsilon_{tsk}(J^{\rm pn}R_{qst0}+J^{\rm pu}R_{qs0t}),
\\[1.5ex]
   \partial_t R_{q0k0}&=&\varepsilon_{isq}h^{\re}_iR_{s0k0}
   +\varepsilon_{isk}h^{\rm n}_iR_{q0s0}+\label{eq:A.6}
 J^{\rm en}\varepsilon_{ksq}(R_{s000}-R_{00s0})\nonumber \\
 &&{}+\varepsilon_{tsq}(J^{\rm ep}R_{stk0}+J^{\rm eu}R_{s0kt})+
 \varepsilon_{stk}(J^{\rm pn}R_{qst0}+J^{\rm nu}R_{q0ts}),
\\[1.5ex]
  \partial_t R_{q00k}&=&\varepsilon_{isq}h^{\re}_iR_{s00k}+\varepsilon_{isk}h^{\rm u}_iR_{q00s}+\label{eq:A.7}
 J^{\rm eu}\varepsilon_{ksq}(R_{s000}-R_{000s})\nonumber \\
  &&{}+\varepsilon_{tsq}(J^{\rm ep}R_{st0k}+J^{\rm en}R_{s0tk})+
 \varepsilon_{tsk}(J^{\rm pu}R_{qt0s}+J^{\rm nu}R_{q0ts}),
\\[1.5ex]
\partial_t R_{0qk0}&=&\varepsilon_{isq}h^{\rm p}_iR_{0sk0}+\varepsilon_{isk}h^{\rm n}_iR_{0qs0}+\label{eq:A.8}
 J^{\rm pn}\varepsilon_{ksq}(R_{0s00}-R_{00s0})\nonumber \\
  &&{}+\varepsilon_{tsq}(J^{\rm ep}R_{tsk0}+J^{\rm pu}R_{0skt})+
 \varepsilon_{stk}(J^{\rm en}R_{sqt0}+J^{\rm nu}R_{0qts}),
\\[1.5ex]
\partial_t R_{0q0k}&=&\varepsilon_{isq}h^{\rm p}_iR_{0s0k}+\varepsilon_{isk}h^{\rm u}_iR_{0q0s}+\label{eq:A.9}
 J^{\rm pu}\varepsilon_{ksq}(R_{0s00}-R_{000s})\nonumber \\
&&{}+\varepsilon_{stq}(J^{\rm ep}R_{st0k}+J^{\rm pn}R_{0tsk})+
 \varepsilon_{stk}(J^{\rm eu}R_{sq0t}+J^{\rm nu}R_{0qst}),
\\[1.5ex]
\partial_t R_{00qk}&=&\varepsilon_{isq}h^{\rm n}_iR_{00sk}+\varepsilon_{isk}h^{\rm u}_iR_{00qs}+\label{eq:A.10}
 J^{\rm nu}\varepsilon_{ksq}(R_{00s0}-R_{000s})\nonumber \\
 &&{}+\varepsilon_{stq}(J^{\rm en}R_{s0tk}+J^{\rm pn}R_{0stk})+
 \varepsilon_{stk}(J^{\rm eu}R_{s0qt}+J^{\rm pu}R_{0sqt}),
\\[1.5ex]
  \partial_tR_{qkl0}&=&\varepsilon_{isq}h^{\re}_iR_{skl0}+\varepsilon_{isk}h^{\rm p}_iR_{qsl0}+\varepsilon_{isl}h^{\rm n}_iR_{qks0}\nonumber
\label{eq:A.11}\\
 &&{}+J^{\rm ep}\varepsilon_{ksq}(R_{s0l0}-R_{0sl0})+J^{\rm pn}\varepsilon_{lsk}(R_{qs00}-R_{q0s0})\nonumber
 \\
  &&{}+J^{\rm en}\varepsilon_{lsq}(R_{sk00}-R_{0ks0})+
 J^{\rm eu}\varepsilon_{stq}R_{tkls}+J^{\rm pu}\varepsilon_{tsk}R_{qslt}+J^{\rm nu}\varepsilon_{tsl}R_{qkst}\,,\qquad
\\[1.5ex]
 \partial_t R_{qk0l}&=&\varepsilon_{isq}h^{\re}_iR_{sk0l}+\varepsilon_{isk}h^{\rm p}_iR_{qs0l}+\varepsilon_{isl}h^{\rm u}_iR_{qk0s}\nonumber
 \label{eq:A.12}\\
 &&{}+J^{\rm ep}\varepsilon_{ksq}(R_{s00l}-R_{0s0l})+J^{\rm pu}\varepsilon_{lsk}(R_{qs00}-R_{q00s})\nonumber
 \\
 &&{}+J^{\rm eu}\varepsilon_{lsq}(R_{sk00}-R_{0k0s})+
 J^{\rm en}\varepsilon_{stq}R_{sktl}+J^{\rm pu}\varepsilon_{tsk}R_{qstl}+J^{\rm nu}\varepsilon_{stl}R_{qkst}\,,
\\[1.5ex]
 \partial_t R_{q0kl}&=&\varepsilon_{isq}h^{\re}_iR_{s0kl}+\varepsilon_{isk}h^{\rm n}_iR_{q0sl}+\varepsilon_{isl}h^{\rm u}_iR_{q0ks}\nonumber
 \label{eq:A.13}\\
 &&{}+J^{\rm en}\varepsilon_{ksq}(R_{s00l}-R_{00sl})+J^{\rm eu}\varepsilon_{lsq}(R_{sok0}-R_{00ks})\nonumber
 \\
 &&{}+J^{\rm nu}\varepsilon_{lsk}(R_{q0s0}-R_{q00s})
 +J^{\rm ep}\varepsilon_{tsq}R_{stkl}+J^{\rm pn}\varepsilon_{stk}R_{qstl}+J^{\rm pu}\varepsilon_{stl}R_{qskt}\,,
\\[1.5ex]
 \partial_t R_{0qkl}&=&\varepsilon_{isq}h^{\rm p}_iR_{0skl}+\varepsilon_{isk}h^{\rm n}_iR_{0qsl}+\varepsilon_{isl}h^{\rm u}_iR_{0qks}\nonumber
 \label{eq:A.14}\\
 &&{}+J^{\rm pn}\varepsilon_{ksq}(R_{0s0l}-R_{00sl})+J^{\rm pu}\varepsilon_{lsq}(R_{0sk0}-R_{00ks})\nonumber
 \\
  &&{}+J^{\rm nu}\varepsilon_{lsk}(R_{0qs0}-R_{0q0s})
 +J^{\rm ep}\varepsilon_{tsq}R_{tskl}+J^{\rm en}\varepsilon_{tsk}R_{tqsl}+J^{\rm eu}\varepsilon_{stl}R_{sqkt}\,,
\\[1.5ex]
 \partial_t R_{qklm}&=&\varepsilon_{isq}h^{\re}_iR_{sklm}+\varepsilon_{isk}h^{\rm p}_iR_{qslm}+\varepsilon_{isl}h^{\rm n}_iR_{qksm}+\nonumber \label{eq:A.15}
\varepsilon_{ism}h^{\rm u}_iR_{qkls}\\
&&{}+ J^{\rm ep}\varepsilon_{ksq}(R_{s0lm}-R_{0slm})
  +J^{\rm en}\varepsilon_{lsq}(R_{sk0m}-R_{0ksm})
\nonumber  \\
&&{}+J^{\rm pn}\varepsilon_{lsk}(R_{qs0m}-R_{q0sm})+ J^{\rm
eu}\varepsilon_{msq}(R_{skl0}-R_{0kls}) \nonumber\\&&{}+J^{\rm
pu}\varepsilon_{msk}(R_{qsl0}-R_{q0ls})+J^{\rm
nu}\varepsilon_{msl}(R_{qks0}-R_{qk0s}).
  \end{eqnarray}
Concrete calculations were carried out for the separable
mixed initial state  \textit {Sep-Mix}:
\begin{equation} \label {A.16}
\rho _ {\rm  Sep-Mix} (0) = \rho^{\rm e} \otimes \rho^{\rm p}
\otimes \rho^{\rm n} \otimes \rho^{\rm u},
\end{equation}
where
\begin{equation} \label {A.17}
\rho^i =\frac {1} {2\cosh h^i_3(0)/2T^i (0)}
\exp\left(-\frac{h^i_3(0)}{2T^i (0)}\sigma_3\right),
\end{equation}
is the mixed state of the $i$ particle, $ T^i (0)$  is  the  initial
temperature, $h^i_3(0)$ is the initial field,\;$i = (e,\ p,\ n,\ u)$.

\ukrainianpart

\title{Нерівноважна оборотна динаміка роботи в 4-спіновій системі у магнітному полі}
\author{Є.О. Іванченко}

\address{Інститут теоретичної фізики,
Національний науковий центр ``Харківський фізико-технічний інститут'',  Харків, Україна}

\makeukrtitle

\begin{abstract}
За допомогою декомплексифікації рівняння Ліувіля-Неймана одержана замкнута система рівнянь локальних векторів Блоха та спінових кореляційних функцій для системи 4 магнітних частинок з обмінною взаємодією, які знаходяться в довільному залежному від часу магнітному полі. Виконано аналітичний та чисельний аналіз динаміки квантових термодинамічних параметрів в залежності від сепарабельного початкового  стану системи та модуляції магнітного поля. В умовах унітарної  еволюції  досліджено  нерівноважну оборотну генерацію роботи у найближчому оточенні.

\keywords  оборотна динаміка, генерація роботи, спінова система, магнітне поле

\end{abstract}

\end{document}